# Elastic properties and inter-atomic bonding in new superconductor $KFe_2Se_2$ from first principles calculations

I.R. Shein *, A.L. Ivanovskii

*Institute of Solid State Chemistry, Ural Branch of the Russian Academy of Sciences, 620990, Ekaterinburg, Russia*

ABSTRACT

Very recently (November, 2010, *PRB*, **82**, *180520R*) the first 122-like ternary superconductor $K_xFe_2Se_2$ with enhanced $T_C$ ~ 31K has been discovered. This finding has stimulated much activity in search of related materials and triggered the intense studies of their properties. Indeed already in 2010-2011 the superconductivity ($T_C$ ~ 27-33K) was also found in the series of new synthesized 122 phases such as $Cs_xFe_2Se_2$, $Rb_xFe_2Se_2$, $(TlK)_xFe_ySe_2$ *etc*. which have formed today the new family of superconducting iron-based materials without toxic As. Here, using the *ab initio* FLAPW-GGA method we have predicted for the first time the elastic properties for $KFe_2Se_2$ and discussed their interplay with inter-atomic bonding for this system. Our data reveal that the examined phase is relatively soft material. In addition, this system is mechanically stable, adopts considerable elastic anisotropy, and demonstrates brittleness. These conclusions agree with the bonding picture for $KFe_2Se_2$, where the inter-atomic bonding is highly anisotropic and includes ionic, covalent and metallic contributions.

* Corresponding author.
*E-mail address:* shein@ihim.uran.ru (I.R. Shein)

# 1. Introduction.

Very recently the first 122-like ternary superconductor $K_xFe_2Se_2$ with enhanced $T_C \sim 31K$ has been discovered [1]. Very quickly a set of others related superconducting phases such as $Cs_xFe_2Se_2$, $Rb_xFe_2Se_2$, $(TlK)_xFe_ySe_2$ *etc*. were successfully synthesized [2-5]. These systems have formed the new group of superconducting iron-based materials without toxic As.

The discovery of these materials has stimulated much activity in various areas of physics and materials science. So, the attention was paid to the transport properties of new 122-like FeSe phases [1-7]. The intense researches of their electronic band structure and magnetic ordering [8-13], of the effects of pressure [14-16] and non-stoichiometry [1-7,11-16] and so on are carried out now.

In this context, the elastic properties of 122-like FeSe phases seem important for various possible applications of these new superconducting materials, as well as can provide some hints on the difference or common features between well known 122 iron-pnictides (reviews [17-21]) and these new iron-selenide systems, and further shed light on the mechanism of superconductivity in Fe-based superconductors. Indeed, the elastic constants can be linked to such important physical parameters of superconductors (SCs) as the Debye temperature $\Theta_D$ and the electron-phonon coupling constant $\lambda$ [22]. Some others correlations between the superconducting critical temperature $T_C$ and mechanical parameters have been also discussed [23].

Motivated by these circumstances, we performed first-principles calculations to predict the elastic properties for $KFe_2Se_2$ as the parent phase of new family of recently discovered 122-like FeSe SCs, and discussed these data in relationship with peculiarities of the inter-atomic bonding in this material.

# 2. Model and computation details.

The new potassium intercalated iron selenide $KFe_2Se_2$ adopts [1] tetragonal $ThCr_2Si_2$-type structure (space group I4/mmm; #139). This structure may be described as the sequence (along the *c* axis) of [$Fe_2Se_2$] blocks separated by K planar sheets: …K/[$Fe_2Se_2$]/K/[$Fe_2Se_2$]/K… The atomic positions are K: 2*a* (0, 0, 0), Fe: 4*d* (½, 0, ½) and Se: 4*e* (0, 0, $z_{Se}$); where $z_{Se}$ is the so-called internal coordinate.

Our calculations were carried out by means of the full-potential method with mixed basis APW+lo (LAPW) implemented in the WIEN2k suite of programs [25]. The generalized gradient approximation (GGA) to exchange-correlation potential in the PBE form [26] was used. The plane-wave expansion was taken up to $R_{MT} \times K_{MAX}$ equal to 8, and the *k* sampling with 12×12×12 *k*-points in the Brillouin zone was used. The calculations were performed with full-lattice optimizations including internal parameter $z_{Se}$. The self-consistent calculations were considered to be converged when the difference in the total energy of the crystal did not exceed 0.01 mRy as calculated at consecutive steps.

## 3. Results and discussion

Firstly, the equilibrium lattice parameters ($a$ = 3.8608 Å, and $c$ =13.8369 Å) for $KFe_2Se_2$ were calculated and are in reasonable agreement with available data [1]. Next, the values of six independent elastic constants ($C_{ij}$) for $KFe_2Se_2$ were evaluated by calculating the stress tensors on different deformations applied to the equilibrium lattice of the tetragonal unit cell, whereupon the dependence between the resulting energy change and the deformation was determined. All these constants ($C_{11}$ = 146 GPa, $C_{12}$ = 57 GPa, $C_{13}$ = 41 GPa, $C_{33}$ = 64 GPa, $C_{44}$ = 32 GPa, and $C_{66}$ = 68 GPa) are positive and satisfy the well-known Born's criteria [27] for mechanically stable tetragonal crystals: $C_{11} > 0$, $C_{33} > 0$, $C_{44} > 0$, $C_{66} > 0$, $(C_{11} - C_{12}) > 0$, $(C_{11} + C_{33} - 2C_{13}) > 0$ and $\{2(C_{11} + C_{12}) + C_{33} + 4C_{13}\} > 0$.

Next, we see that $C_{33}$ is much smaller than $C_{11}$. This indicates that this layered crystal should be more compressible along the $c$-axis. Besides, the $C_{44}$ constants are lower than $C_{66}$, indicating that the shear along the (100) plane is easier relative to the shear along the (001) plane.

The calculated constants $C_{ij}$ allow us to obtain the macroscopic elastic parameters for $KFe_2Se_2$, namely the bulk ($B$) and the shear ($G$) modules. These values as calculated in two main approximations: Voigt (V) [28] and Reuss (R) [29] as $B_V$ = 1/9$\{2(C_{11} + C_{12}) + C_{33} + 4C_{13}\}$; $G_V$ = 1/30($M$ + 3$C_{11}$ - 3$C_{12}$ + 12$C_{44}$ + 6$C_{66}$); $B_R = C^2/M$; $G_R = 15\{18B_V/C^2 + 6/(C_{11} - C_{12}) + 6/C_{44} + 3/C_{66}\}^{-1}$, where $C^2 = (C_{11} + C_{12})C_{33} - 2C_{13}^2$ and $M = C_{11} + C_{12} + 2C_{33} - 4C_{13}$. The obtained values are: $B_V$ = 70.4 GPa, $G_V$ = 40.9 GPa, and $B_R$ = 57.7 GPa, $G_R$ = 34.9 GPa. Further, within the Voigt-Reuss-Hill (VRH) approximation [30], it is possible to estimate the corresponding effective modules for polycrystalline $KFe_2Se_2$ as the arithmetic mean of these two limits: $B_{VRH}$ = 64 GPa, $G_{VRH}$ = 37.9 GPa, as well as the averaged compressibility $\beta_{VRH} = 1/B_{VRH} = 0.01563$ GPa$^{-1}$, Young modulus $Y_{VRH} = 9B_{VRH}G_{VRH}/(3B_{VRH} + G_{VRH})$ = 94.9 GPa, and the Poisson's ratio $v = (3B_{VRH} - 2G_{VRH})/\{2(3B_{VRH} + G_{VRH})\}$ = 0.253. Also, the Lame's constants (physically, the first constant $\lambda$ represents the compressibility of a material while the second constant $\mu$ reflects its shear stiffness) were estimated as $\mu = Y_{VRH}/2(1+v)$ = 37.8, and $\lambda = vY_{VRH}/\{(1+v)(1-2v)\}$ = 38.8.

Thus, our data reveal that the examined phase is a relatively soft material. Let us note also that the elastic modules of $KFe_2Se_2$ are comparable as a whole with the same for others layered FeAs SC's. For example, according to the available experimental and theoretical data, their bulk moduli are: for $SrFe_2As_2$ $B \sim 62$ GPa, for $CaFe_2As_2$ $B \sim 60$ GPa for for LiFeAs $B \sim 57$ GPa and $B \sim 100 \div 120$ GPa for some of 1111 FeAs phases such as LaFeAsO or NdFeAsO [31-37].

The calculated elastic parameters allow us to make the following conclusions. For $KFe_2Se_2$ $B > G$; this implies that the parameter limiting the mechanical stability of this system is the shear modulus. Since the shear modulus *G* represents the resistance to plastic deformation while the bulk modulus *B* represents the resistance to fracture, the value *B/G* was proposed [38] as a empirical malleability measure of polycrystalline materials: if *B/G* > 1.75, a material behaves in a ductile manner, and *vice versa*, if *B/G* < 1.75, a material will demonstrates brittleness. In our case, *B/G* ~ 1.69, thus this phase should be brittle.

The values of the Poisson ratio (*v*) for covalent materials are small (*v* = 0.1), whereas for ionic materials a typical value is 0.25 [39]. In our case *v* = 0.253, *i.e.* a considerable ionic contribution in inter-atomic bonding for this phase should be assumed. Besides, for covalent and ionic materials, typical relations between bulk and shear moduli are *G* ~ 1.1*B* and *G* ~ 0.6*B*, respectively [39]. In our case, the calculated value of *G/B* is 0.59.

The elastic anisotropy of crystals is an important parameter for engineering science since it correlates with the possibility of emergence of microcracks in materials. Some ways are used to estimate the elastic anisotropy of crystals. So, elastic anisotropy of the tetragonal KFe$_2$Se$_2$ may be estimated by coefficients *A* (note that for crystals with isotropic elastic properties *A* = 1, while values smaller or greater than unity measure the degree of elastic anisotropy see, for example [40]) calculated for every symmetry plane as: $A^{(010),(100)} = C_{44}(C_{11} + 2C_{13} + C_{33})/(C_{11}C_{33} + C_{13}^2)$, and $A^{(001)} = 2C_{66}(C_{11} - C_{12})$. Another way implies the estimation of elastic anisotropy (in percents) in compressibility ($A_B$) and shear ($A_G$), using a model [41] for polycrystalline materials as: $A_B = (B_V - B_R)/(B_V + B_R)$, and $A_G = (G_V - G_R)/(G_V + G_R)$. Finally, the so-called universal anisotropy index [42] defined as: $A^U = 5G_V/G_R + B_V/B_R - 6$ was used; for isotropic crystals $A^U = 0$; deviations of $A^U$ from zero define the extent of crystal anisotropy. The following values of the aforementioned indexes for KFe$_2$Se$_2$ were obtained: $A^{(010),(100)} = 0.85$; $A^{(001)} = 1.53$; $A_B = 9.9\%$, $A_G = 7.9\%$, and $A^U = 1.08$, indicating considerable elastic anisotropy for this layered material.

Thus, using the above calculated elastic parameters, we made the following preliminary conclusions about the bonding picture in the tetragonal KFe$_2$Se$_2$: (i). a considerable ionic contribution in inter-atomic bonding should be assumed for this system; (ii). the atomic bonds along the *c*-axis should be weaker than those along the *a, b* axis, and (iii). the inter-blocks (..[Fe$_2$Se$_2$]/K/[Fe$_2$Se$_2$]..) bonding and the bonding inside of [Fe$_2$Se$_2$] blocks should be different enough to provide the above elastic anisotropy.

For the further detailed understanding of this picture, we will address to the results of electronic structure calculations of KFe$_2$Se$_2$. The calculated total and atomic-resolved partial densities of states (DOSs) are depicted in Fig. 1. We see that the states between -2.7 eV and the Fermi level ($E_F$ = 0 eV) are mainly of the Fe 3*d* character, forming metallic-like Fe-Fe bond. In turn, the Se 4*p* - Fe 3*d* states, which are placed between -6.5 eV and -3.7 eV with respect to the Fermi level, are strongly hybridized, and are responsible for the covalent bonding Fe-Se inside [Fe$_2$Se$_2$] blocks. These covalent bonds are clearly visible in Fig. 2, where the charge density map for KFe$_2$Se$_2$ is depicted. It is seen also that weak direct Se-Se inter-block interactions occur.

Within simplified ionic model if we use the usual oxidation numbers of atoms ($K^{1+}$, $Fe^{2+}$, and $Se^{2-}$), the charge transfer ($1e$) from $K^{1+}$ sheets to $[Fe_2Se_2]^{1-}$ blocks should be assumed. To estimate numerically the amount of electrons, redistributed between the various atoms as well as between adjacent blocks, we carried out a Bader [42] analysis, and the total charges of atoms - the so-called Bader charges, $Q^B(K) = 8.22e$, $Q^B(Fe) = 7.50e$, and $Q^B(Se) = 6.89e$ were obtained. Next, the corresponding effective charges (as obtained from the purely ionic model ($Q^i$) and their differences ($\Delta Q = Q^B - Q^i$)) are: $\Delta Q(K) = +0.78e$, $\Delta Q(Fe) = +0.50e$, and $\Delta Q(Se) = -0.89e$. The results confirm that $KFe_2Se_2$ is partly ionic compound, and the charges are transferred from K and Fe to Se. Thus the charge transfer from $K^{0.78+}$ sheets to $[Fe_2Se_2]^{0.78-}$ blocks takes place. Summarizing these results, the picture of chemical bonding for the $KFe_2Se_2$ phase may be described as a high-anisotropic mixture of ionic, covalent and metallic contributions, where (i). inside $[Fe_2Se_2]$ blocks, mixed covalent-ionic bonds Fe-Se take place owing to hybridization of Fe $3d$ – Se $4p$ states and Fe $\rightarrow$ Se charge transfer; in addition metallic-like Fe-Fe bonds appear owing to delocalized near-Fermi Fe $3d$ states; (ii) between adjacent $[Fe_2Se_2]$ blocks and potassium sheets, ionic bonds emerge owing to K $\rightarrow$ $[Fe_2Se_2]$ charge transfer, and (iii) between the adjacent $[Fe_2Se_2]/[Fe_2Se_2]$ blocks, weak covalent Se-Se bonds occur as a result of hybridization of $p$-$p$ states of Se atoms.

## 4. Conclusions.

In summary, we have for the first time determined the set of the main elastic parameters (such as independent elastic constants, bulk and shear modules, Young's modulus, Poisson's ratio, indicators of elastic anisotropy and brittle/ductile behavior *etc.*) for $KFe_2Se_2$ - as the parent phase of new family of very recently (2010-2011) discovered 122-like FeSe superconducting materials, and have analyzed their interplay with inter-atomic bonding for this system.

Our studies showed that $KFe_2Se_2$ is a mechanically stable and is relatively soft material ($\beta \sim 0.0156$ GPa$^{-1}$), adopts considerable elastic anisotropy and will demonstrates brittleness. Our data reveal that such elastic behavior of $KFe_2Se_2$ agrees with the bonding picture, according to which the inter-atomic bonding in this system is highly anisotropic and includes ionic, covalent and metallic contributions, where inside $[Fe_2Se_2]$ blocks the mixed covalent-ionic Fe-Se and metallic-like Fe-Fe bonds take place, whereas inter-blocks (..K/$[Fe_2Se_2]$/K/$[Fe_2Se_2]$..) bonding is basically of ionic type (owing to K $\rightarrow$ $[Fe_2Se_2]$ charge transfer) with additional weak covalent Se-Se bonding.

**Acknowledgement**

Financial support from the RFBR (grants No. 09-03-00946, and No. 10-03-96008) is gratefully acknowledged.

**FIGURES**

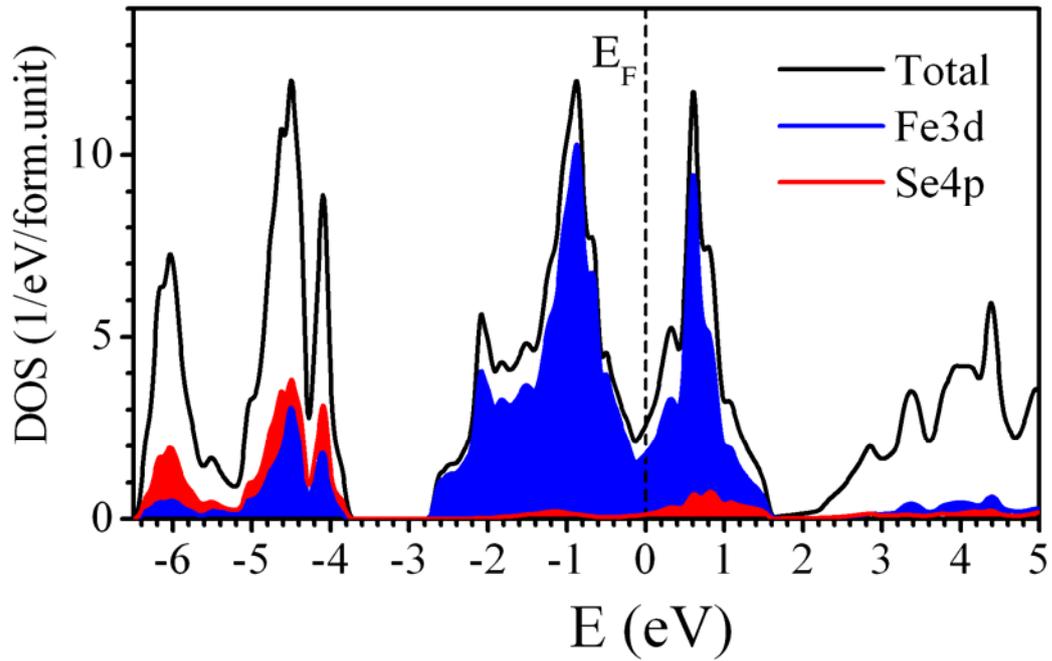

**Figure 1.** Total and partial densities of states for KFe$_2$Se$_2$.

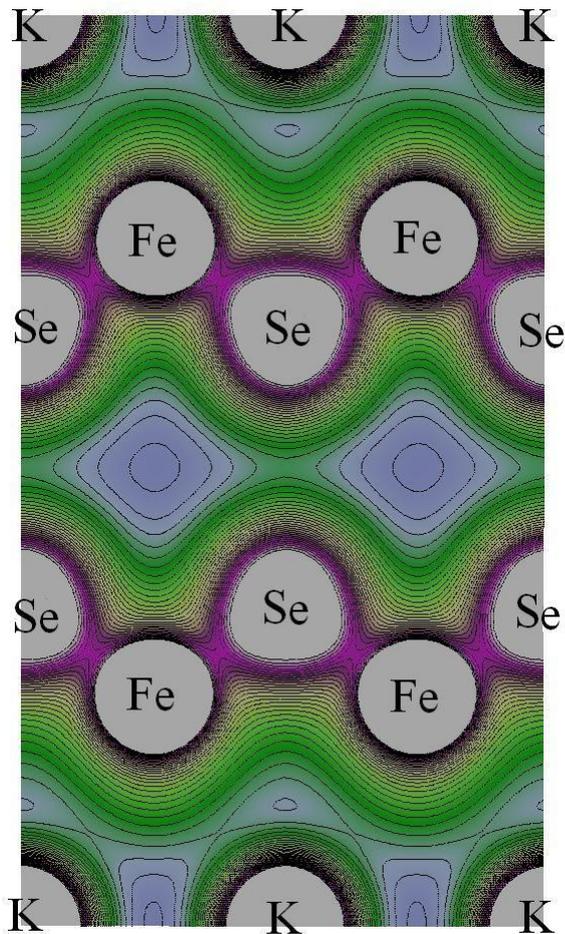

**Figure 2.** Charge density map in (001) plane for KFe$_2$Se$_2$.

8